\pdfoutput=1 
\documentclass{article}
\PassOptionsToPackage{numbers}{natbib}

\usepackage[preprint]{neurips_2019}

\usepackage[utf8]{inputenc} 
\usepackage[T1]{fontenc}    
\usepackage{hyperref}       
\usepackage{url}            
\usepackage{booktabs}       
\usepackage{amsfonts}       
\usepackage{nicefrac}       
\usepackage{microtype}      

\usepackage{times}
\usepackage{epsfig}
\usepackage{graphicx}
\usepackage{amsmath}
\usepackage{amssymb}
\usepackage{xcolor}
\usepackage{siunitx}
\usepackage{caption}
\usepackage{float}
\usepackage{booktabs}

\usepackage{bbm}
\usepackage{amsthm}
\usepackage{subcaption}
\usepackage{color, colortbl}

\definecolor{Gray}{gray}{0.9}
\definecolor{LightCyan}{rgb}{0.9,0.95,1}
\definecolor{LightRed}{rgb}{1,0.95,0.95}

\title{Self-supervised audio representation learning \\ for mobile devices}

\author{%
  Marco Tagliasacchi \\
  Google Research\\
  \texttt{mtagliasacchi@google.com} \\
  \And
  Beat Gfeller \\
  Google Research\\
  \texttt{beatg@google.com} \\
  \And
  F\'elix de Chaumont Quitry\\
  Google Research\\
  \texttt{fcq@google.com} \\
  \And
  Dominik Roblek\\
  Google Research\\
  \texttt{droblek@google.com} \\
}

\newcommand{\audiotovec}{\texttt{Audio2Vec}} 
\newcommand{\autoencoder}{\texttt{AutoEncoder}} 
\newcommand{\temporalgap}{\texttt{TemporalGap}} 
\newcommand{\tripletloss}{\texttt{TripletLoss}} 
 
\newcommand{\wordtovec}{\texttt{Word2Vec}} 
\newcommand{\speechtovec}{\texttt{Speech2Vec}} 

\newcommand{\spectrogram}{\texttt{Spectrogram}} 
\newcommand{\untrained}{\texttt{Untrained}}

\newcommand{\multihead}{\texttt{MultiHead}}
\newcommand{\singlehead}{\texttt{Supervised}}

\newcommand{\audioset}{\emph{AudioSet}} 
\newcommand{\speechcommands}{\emph{Speech Commands}} 
\newcommand{\librispeech}{\emph{LibriSpeech}} 
\newcommand{\TUT}{\emph{TUT Urban Acoustic Scenes 2018}} 
\newcommand{\birdsong}{\emph{Bird Audio Detection}}
\newcommand{\MUSAN}{\emph{MUSAN}}
\newcommand{\langid}{\emph{Spoken Language Identification}}

\newcommand{\Dim}{d}

\begin{document}

\maketitle
\begin{abstract}
We explore self-supervised models that can be potentially deployed on mobile devices to learn general purpose audio representations. Specifically, we propose methods that exploit the temporal context in the spectrogram domain. One method estimates the temporal gap between two short audio segments extracted at random from the same audio clip. The other methods are inspired by {\wordtovec}, a popular technique used to learn word embeddings, and aim at reconstructing a temporal spectrogram slice from past and future slices or, alternatively, at reconstructing the context of surrounding slices from the current slice. We focus our evaluation on small encoder architectures, which can be potentially run on mobile devices during both inference (re-using a common learned representation across multiple downstream tasks) and training (capturing the true data distribution without compromising users' privacy when combined with federated learning). We evaluate the quality of the embeddings produced by the self-supervised learning models, and show that they can be re-used for a variety of downstream tasks, and for some tasks even approach the performance of fully supervised models of similar size.
\end{abstract}

\section{Introduction}\label{sec:intro} 
Thanks to advances in supervised audio learning, it is now possible to train models that are able to successfully perform different tasks, including audio annotation~\citep{Hershey2017a}, music recognition~\cite{Arcas2017}, automatic speech recognition~\cite{Chan2016}, speaker identification~\cite{Matejka2016}, etc. Such supervised models can also be deployed on mobile devices by applying network pruning and quantization techniques~\cite{Howard2017, Sandler2018, Frankle2019}.

Despite the indisputable success, this approach suffers from three main shortcomings. First, it requires collecting large annotated datasets specific to each task to be solved. Second, separate models are typically trained for each task, making it difficult to reuse computational resources when multiple such models are deployed on a mobile device. Third, inference is performed on device, but model training is still done on the server side using datasets representing surrogate distributions, which might potentially differ from the true data distribution. 

Unsupervised learning attempts to overcome these limitations, by making it possible to learn from widely available unlabelled datasets and by learning general purpose representations that can be reused for different downstream tasks. In addition, unsupervised learning lends itself to be deployed on device, where no explicit labeling of the data is available. Therefore, by leveraging the recent advances in federated learning~\cite{Bonawitz2019}, it might be possible to distribute the training process across numerous devices, thus training models directly on the true data distribution, while fully preserving users' privacy.

In the area of unsupervised learning, self-supervised learning has emerged as an attractive approach. In a nutshell, an auxiliary task is formulated based on the available unlabelled data and a fully supervised model is trained to solve such a task. The key idea is that, by solving the auxiliary task, the model is also learning some general purpose representations in a lower dimensional embedding space. Therefore, the embedding encoder, e.g., the portion of the model architecture mapping the input data to the embedding space, can be reused as a feature extractor for different downstream tasks. 

\begin{figure}
  [t] \center{
  \includegraphics[width=1.0\textwidth]{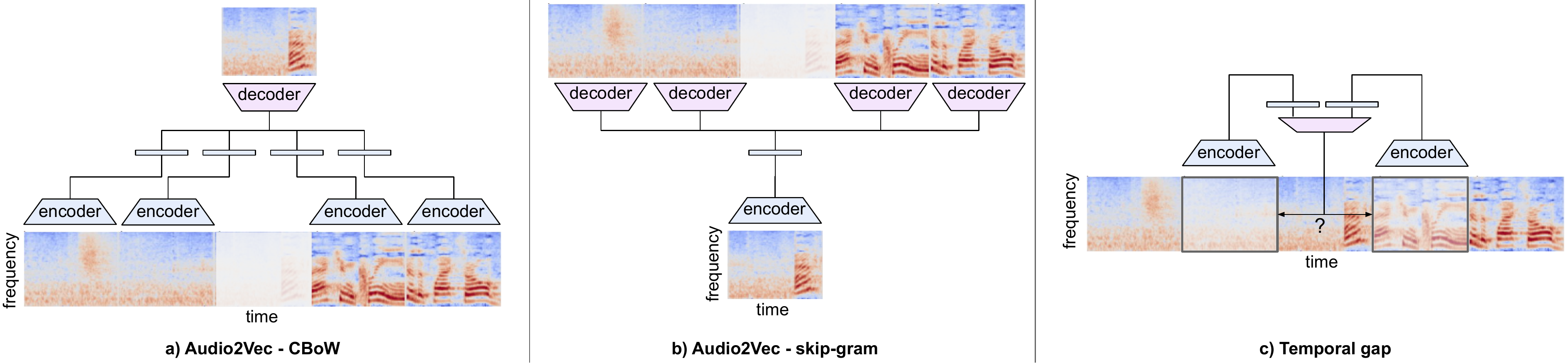}} 
  \caption{\label{fig:self_tasks} Overview of the proposed self-supervised learning tasks.} 
\end{figure}

One of the earliest successes of self-supervised learning was obtained in the context of language models, where {\wordtovec} is used to map one-hot-encoded words to word embeddings~\cite{Mikolov2013a}. {\wordtovec} can be formulated in two variants: i) continuous bag-of-words (CBoW), or ii) skip-gram. In the former, the model predicts the current word based on the context of surrounding words. In the latter, the model predicts surrounding words given the current word.
Recently, a similar approach has been proposed to map speech to fixed-dimensional embeddings~\cite{Chung2018b, Chung2018c}. The {\speechtovec} architecture consists of a RNN encoder-decoder which can handle variable-length inputs and outputs.

In this paper we explore self-supervised learning of audio representations, using a small model architecture which can be potentially deployed on mobile devices during both training and inference. We posit that contextual temporal information can be exploited in the case of general audio signals without resorting to any form of explicit supervision. We argue that solving properly designed tasks that involve the temporal context requires extracting some sort of high level semantic information from the underlying raw data, thus leading to reusable embeddings. In this respect, this paper makes the following main contributions:
\begin{itemize}
  \item We propose {\audiotovec}, a self-supervised learning task that is inspired by {\wordtovec}, but applied to audio spectrograms. In the CBoW formulation (Figure~\ref{fig:self_tasks}a) the auxiliary task consists of reconstructing a temporal slice of pre-determined duration from a number of past and future slices. In the skip-gram formulation (Figure~\ref{fig:self_tasks}b) the roles of the target and surrounding slices are reversed.
  
  \item We propose {\temporalgap}, a self-supervised learning task that consists of estimating the distance in time between any two pairs of audio segments extracted at random from a longer audio clip (Figure~\ref{fig:self_tasks}c).
  
  \item We quantitatively evaluate the quality of the embeddings produced by the feature encoders obtained by solving the aforementioned self-supervised tasks. To this end we consider a wide variety of downstream tasks, including speech, music detection, speaker identification and language identification, among others. Our results show that all self-supervised models are able to partially bridge the accuracy gap with fully supervised models.
  
  \item We focus our evaluation on small encoder architectures, which can be suitably deployed on mobile devices~\cite{Howard2017, Sandler2018, Frankle2019}. During inference, it makes it possible to explore the potential offered by self-supervised learning to share computational resources across different tasks, by using a common embedding encoder. During training, it enables to capture the true data distribution when used together with federated learning~\cite{Bonawitz2019}.
\end{itemize}

The rest of this paper is organized as follows. Section~\ref{sec:related_work} discusses the most relevant literature related to our work. Section~\ref{sec:method} presents the proposed methods, which are evaluated in Section~\ref{sec:experiments}. Conclusions and future work are given in Section~\ref{sec:conclusions}.

\section{Related work}\label{sec:related_work}
The work presented in this paper is related to several different areas that have received attention in the recent literature. In particular, learning representations has been explored for different modalities.

\textbf{Learning audio representations}:
Unsupervised feature learning can lead to more compact and descriptive representations than traditional handcrafted features, e.g., MFCCs.
For example,~\cite{Lee2009} adopt convolutional deep belief networks to learn audio representations, applicable to both speech and music related tasks. More recently, different autoencoder architectures have been explored, e.g., denoising~\cite{Xu2017}, convolutional LSTM autoencoders~\cite{Meyer2017} and sequence-to-sequence autoencoders~\cite{Chung2016}. A self-supervised version of the triplet loss is proposed in~\cite{Jansen2018}. In the absence of labels, the authors create anchor-positive pairs by adding noise, shifting in time and/or frequency, and sampling temporal neighbors. When tested on AudioSet~\cite{Gemmeke2017}, the self-supervised embeddings partially bridge the gap between a simple log spectrogram baseline and a fully supervised classifier.

\textbf{Learning visual representations}:
Several auxiliary tasks have been explored to learn image representations, e.g., predicting the relative position of a pair of patches extracted from the same image~\cite{Doersch2015}, re-ordering image patches and solving jigsaw puzzles~\cite{Noroozi2016}, or asking the model to discriminate between a patch and transformed version of it~\cite{Dosovitskiy2016}.
In some cases solving seemingly simple tasks can lead to very powerful representations such as, for example, detecting image rotations~\cite{Gidaris2018}. In other cases, representations can be learned as a by-product of solving useful tasks, e.g., in the case of image colorization~\cite{Zhang2016} and image inpainting~\cite{Pathak2016}. The latter is to some extent similar to our work, since the CBoW version of {\audiotovec} can be seen as a form of inpainting in the spectrogram domain. The representations learned by different self-supervised learning tasks can also be combined to obtain a single representation, as presented in~\cite{Doersch2017}.
In the case of video, it is possible to exploit the temporal dimension to learn visual representations by asking a model to learn whether frames are in the correct temporal order~\cite{Misra2016, Fernando2017}, to infer motion by observing a static image~\cite{Pathak2017}, or detect whether a video is playing forwards or backwards~\cite{Wei2018}.

\textbf{Learning multimodal representations}:
Several papers have recently investigated learning audio representations exploiting the correlation with other modalities, e.g., text~\cite{Chung2018a}, images~\cite{Owens2018b} and videos~\cite{Owens2018a, Gao2018, Arandjelovic2017, Korbar2018,Cramer2019}.

\textbf{Contextual predictions}:
After the seminal work on {\wordtovec}, contextual prediction has been successfully explored as a means for learning representations in other modalities, e.g., in the case of image inpainting~\cite{Pathak2016}, symbolic music prediction~\cite{Bretan2017}, and speech~\cite{Chung2018b}.  Recently \cite{VandenOord2019} proposed to use contrastive predictive coding, i.e., predicting future samples directly in the embedding space, reporting promising results also in the case of audio-based tasks. Our work is mostly related to this strand of research, in that we evaluate contextual prediction for general audio-based tasks, but we put particular emphasis on learning models that can be deployed on device.

\section{Methods}\label{sec:method}

\newcommand{\x}{x}
\newcommand{\X}{X}
\newcommand{\Xh}{\hat{X}}
\newcommand{\z}{z}
\newcommand{\Enc}{Enc}
\newcommand{\Dec}{Dec}

Let $\x = \{x_1, x_2, \ldots, x_{n}\}$ denote an audio clip of $n$ samples in the time domain and $\X \in \mathbb{R}^{T \times F}$ the corresponding real-valued log-mel spectrogram, which consists of $T$ temporal frames and $F$ frequency bins.
Let $\X_i$ denote a $N \times F$ slice of the spectrogram $\X$, starting at frame $i$ with $N < T$ temporal frames and $\z_i = \Enc(\X_i)$ a $\Dim$-dimensional embedding computed by processing the input spectrogram $\X_i$ with an encoder $\Enc()$, whose architecture is detailed in Section~\ref{sec:experiments}. Using this notation, in the following we describe the proposed self-supervised learning models. 

\textbf{Audio2Vec (CBoW)}: The first self-supervised learning task that we propose, $\audiotovec$, comes in two variants. In the CBoW variant, we first select a target slice at random, together with a set of surrounding slices used for prediction. Each of the predictor slices is processed by the same encoder, which maps its input into a fixed-dimensional embedding. These embeddings are then concatenated and fed into a decoder, mimicking the same architecture as the encoder, which computes a reconstruction of the target slice. More specifically, let $\X_{(0)} = \X_i$ be a slice selected at random from $\X$. Then, a set of past ($\X_{(-P)}, \ldots, \X_{(-1)}$) and future slices ($\X_{(1)}, \ldots, \X_{(P)}$) are extracted from the same audio clip. The temporal location of the slice $\X_{(p)}$ is equal to $\X_{i + p(N + G)}$, i.e., we consider non-overlapping slices of size $N$, with an extra gap of $G$ temporal frames between any two consecutive slices. The gap is introduced to avoid that the self-supervised model exploits the leakage between adjacent STFT temporal frames as a \emph{shortcut} to solve the task. Each slice is processed by the same encoder to obtain $\z_{(p)} = \Enc(\X_{(p)})$. Then, a vector $\z_{(0)} = [\z_{(-P)}, \ldots, \z_{(-1)}, \z_{(1)}, \ldots, \z_{(P)}]$ is obtained by concatenating the embeddings of each of the predictor slices and fed into a convolutional decoder to obtain a reconstruction $\Xh_{(0)} = \Dec(\z_{(0)})$. Note that the architecture of the decoder is obtained by reversing the order of the layers in the encoder and replacing max-pooling with nearest-neighbor upsampling. The overall encoder-decoder architecture is trained end-to-end by minimizing the mean-square error loss function $\| \X_{(0)} - \Xh_{(0)}\|$. 

\textbf{Audio2Vec (skip-gram)}: The skip-gram variant of $\audiotovec$ uses a similar architecture. In this case we compute the embeddings of the middle slice $\z_{(0)} = \Enc(\X_{(0)})$, and then let the decoder reconstruct the surrounding slices, i.e., $[\Xh_{(-P)}, \ldots, \Xh_{(-1)}, \Xh_{(1)}, \ldots, \Xh_{(P)}] = \Dec(\z_{(0)})$. The decoder is identical to the one used by the CBoW variant, except for one important difference: the last convolutional layer has $2P$ output channels, one for each of the slices to be reconstructed. The loss function minimizes the average mean-square error computed across the $2P$ reconstructed slices.

\textbf{Temporal gap}: For the {\temporalgap} task, we ask the model to estimate the absolute value of the distance in time between two slices sampled at random from the same audio clip. More specifically, we sample the ground truth temporal gap from a uniform distribution, i.e., $\Delta \sim \mathcal{U}(0, N_{max} - N)$, where $N$ and $N_{max}$ are the lengths (in time frames) of the slices and the original sample, respectively, and define the normalized temporal gap as $\delta = \Delta / (N_{max} - N) \in [0, 1]$. Then, we extract two slices $\X_i$ and $\X_j$ such that $\Delta = |i - j|$. Note that we do not impose a temporal order between the two slices. We concatenate the embedding representations in a single $2\Dim$-dimensional vector $\z = [\Enc(\X_i), \Enc(\X_j)]$ and we feed this vector into a fully connected feed
forward network with a single hidden layer of size 64 that produces the scalar output $\hat{\delta}$. We train the model end-to-end so as to minimize a cross-entropy loss $\mathcal{L}_{CE}(\delta, \hat{\delta})$ between the ground-truth and the predicted gap. In our experiments, we found that this loss is to be preferred to the mean-square error $\| \delta - \hat{\delta} \|$, presumably because it gives more weight to errors when the ground truth gap $\delta$ is small.

\section{Experiments}\label{sec:experiments}
We compare the quality of the embeddings produced by different self-supervised learning methods according to two different evaluation measures: i) the accuracy of a fully supervised logistic regression model trained using the embeddings and the corresponding labels as inputs; ii) the accuracy of a non-parametric nearest neighbors model that works directly in the embedding space. In the following we describe the datasets used in our evaluation campaign and the baselines to which we compare our results.

\textbf{Datasets}: We use {\audioset}~\cite{Gemmeke2017} to train all the self-supervised learning tasks, unless stated otherwise. {\audioset} contains excerpts of 10 seconds from the soundtracks of YouTube videos. Although the dataset is annotated with labels of more than 500 classes, we discard them in our study. Note that each AudioSet sample can be potentially reused multiple times during training, each time extracting a different target slice (together with surrounding slices) uniformly at random. 

We use six publicly available datasets to evaluate a variety of downstream tasks, covering both speech and non-speech related tasks. We use the {\speechcommands} dataset~\cite{Warden2018} to evaluate keyword spotting on 35 distinct keywords. 
{\librispeech}~\cite{Panayotov2015} contains audio books read by 251 different speakers. We use the 100 hours training set to evaluate a speaker identification task. The {\langid} dataset~\cite{Tomasz2018} contains samples that belong to three different languages: English, Spanish and German, while the {\MUSAN} dataset~\cite{Snyder2015} distinguishes across three classes, namely music, speech and noise. Finally, we use two datasets released in the context of the recent DCASE2018 Challenge, {\birdsong}~\cite{Stowell2018} and {\TUT}~\cite{Mesaros2018}, which contains labeled audio samples from 10 different urban environments. To the best of the authors' knowledge, this is the first paper that comprehensively evaluates the quality of self-supervised learning for audio on such a wide variety of tasks. 

Since each dataset is characterized by samples having different durations, during training we preprocess the downstream datasets extracting equal-length slices uniformly at random from the original sample and assign the corresponding label to all of the extracted slices. We consider input samples having the duration of $T=975$~ms, so as to match the size of the temporal slices used when training the self-supervised tasks. During evaluation, we apply a sliding window of size $T$ and a hop size of $T/2$, so as to one or more predictions for each input samples, depending on its length. In order to aggregate such predictions and produce a single output for each sample, we apply a simple naive-Bayes classifier.

\textbf{Encoder architecture}: In our work we consistently use the same audio frontend, which processes input sequences sampled at 16~kHz, with a window size of 25~ms and a hop size equal to 10~ms to compute the short-time Fourier transform (STFT), and then computes $F=64$ mel-spaced frequency bins in the range 60--7800~Hz. 
For the encoder $\Enc()$, we use a convolutional neural network, whose architecture is described in Table~\ref{tab:encoder_architecture}. Due to its limited size (approximately 125k parameters) it can be potentially deployed on a mobile device and run in an energy-efficient way by exploiting streaming convolutions. Each convolutional layer consists of a series of two convolutions, one along the time axis (with size $3 \times 1 \times C_{in} \times C_{out}$) and one along the frequency axis (with size $1 \times 3 \times C_{in} \times C_{out}$), in parallel to a pointwise $1 \times 1$ convolution as a residual connection. All activation functions are ReLUs and batch normalization is used in all convolutional layers. Each layer is followed by max-pooling, to reduce the time-frequency dimensions by a factor of two at each layer. Finally, a global max-pooling layer produces a $\Dim$-dimensional vector, which is further processed by a fully-connected layer to get the embeddings. By default, we set $N=96$ (corresponding to 975~ms) and $\Dim=128$, thus reducing the dimensionality of the raw audio samples by a factor of about $122$.

\begin{table}
  \begin{center}
    \caption{Encoder architecture. Size of activations, number of parameters and FLOPs.}
    \label{tab:encoder_architecture}%
    
      \begin{tabular}{|c|c|c|c|} \hline
       & Output Size & Num. params & FLOPs \\ 
       \hline
       Input layer & $96 \times 64 \times 1$ & - & - \\
       Conv. layer 1 & $48 \times 32 \times 8$ & 0.2k & 2.9M  \\ 
       Conv. layer 2 & $24 \times 16 \times 16$ & 1k & 4M \\ 
       Conv. layer 3 & $12 \times 8 \times 32$ & 5k & 4M \\    
       Conv. layer 4 & $6 \times 4 \times 64$ & 20k & 3.9M \\ 
       Conv. layer 5 & $3 \times 2 \times 128$ & 82k & 3.9M \\       
       FC layer & $1 \times 1 \times 128$ & 16k & 33k \\  
       \hline  
   
       Total & - & 125k & 18.7M \\
       \hline
      \end{tabular}
  \end{center}

\end{table}
\textbf{Models}: We compare different self-supervised models trained on {\audioset}: {\audiotovec}, in its two variants, CBoW and skip-gram, and {\temporalgap}. For {\audiotovec} we use $P=2$ slices on each side of the target, and a gap of $G=2$ temporal frames between consecutive slices. 
We also include the {\tripletloss} methods proposed in~\cite{Jansen2018} in our evaluations. More specifically, positive/negative pairs are obtained by extracting a slice from, respectively, the same or a different original sample. In addition, we also train an {\autoencoder} sharing the same encoder and decoder architectures as {\audiotovec}. We tested different variants, including denoising and variational autoencoders, but we did not observe significant differences with respect to the default autoencoder. When evaluating the accuracy in downstream tasks, we extract the portion of the model corresponding to the encoder and use it to map input log-mel spectrograms to $128$-dimensional embeddings.

We compare our results to two different fully supervised baselines based on a simple logistic regression model: i) the {\spectrogram} model receives directly the (flattened) spectrogram features as input; ii) the {\untrained} model computes the embeddings with the very same encoder architecture described in Section~\ref{sec:method}, but using randomly initialized weights.  

Since each task is charaterized by a different number of target classes and intrinsic difficulty, we compare the accuracy to the level attained by task-specific fully supervised models ({\singlehead}), each using the same encoder, but trained end-to-end on each of the labeled downstream datasets. In addition, we also trained a {\multihead} model,
where a single shared encoder is composed with a different fully connected layer for each downstream task. This provides an upper bound for the best performance we could expect, as
it uses the same architecture as when using the self-supervised embeddings, but leverages the in-domain labeled data for end-to-end training.

All models are trained with stochastic gradient descent and Adam optimizer with default hyperparameters. The learning rate was set to $10^{-3}$ for {\audiotovec}, {\autoencoder}, and all the supervised models, while it was set to $10^{-4}$ for {\temporalgap} and {\tripletloss} . We use a mini-batch size equal to 256 and we stop training after approximately 2 days (on five Tesla V100 GPUs), thus iterating between 1.3 and 1.8 million mini-batches. In most cases, the accuracy of downstream tasks saturated after iterating over 500k mini-batches.

\begin{figure}
    \centering
    \begin{subfigure}[b]{0.39\textwidth}        \includegraphics[width=\textwidth]{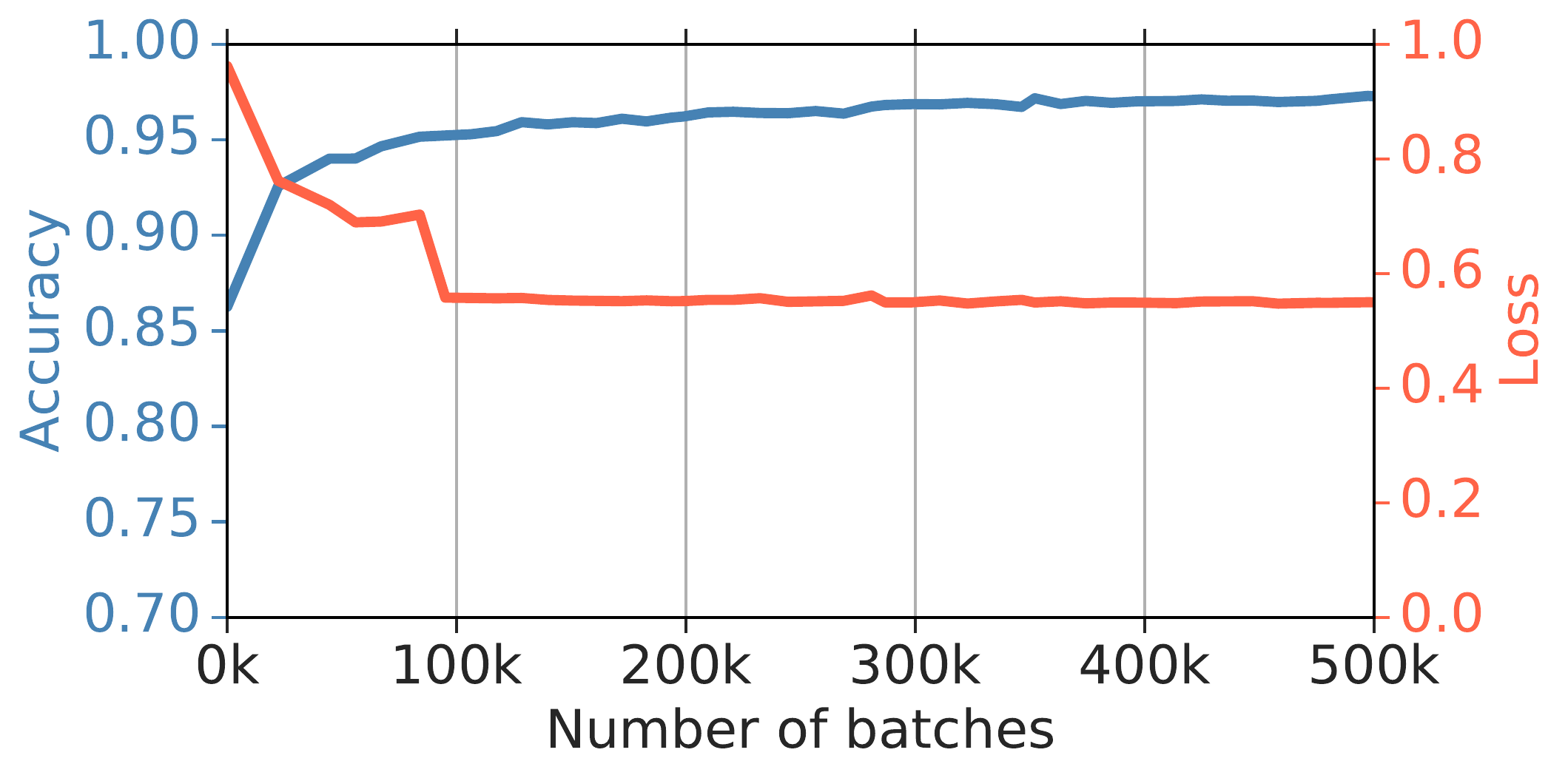}
        \caption{}
        \label{fig:self_task_audio_to_vec_skip_gram_MUSAN}
    \end{subfigure}
    ~ 
    \begin{subfigure}[b]{0.59\textwidth}
        \includegraphics[width=\textwidth]{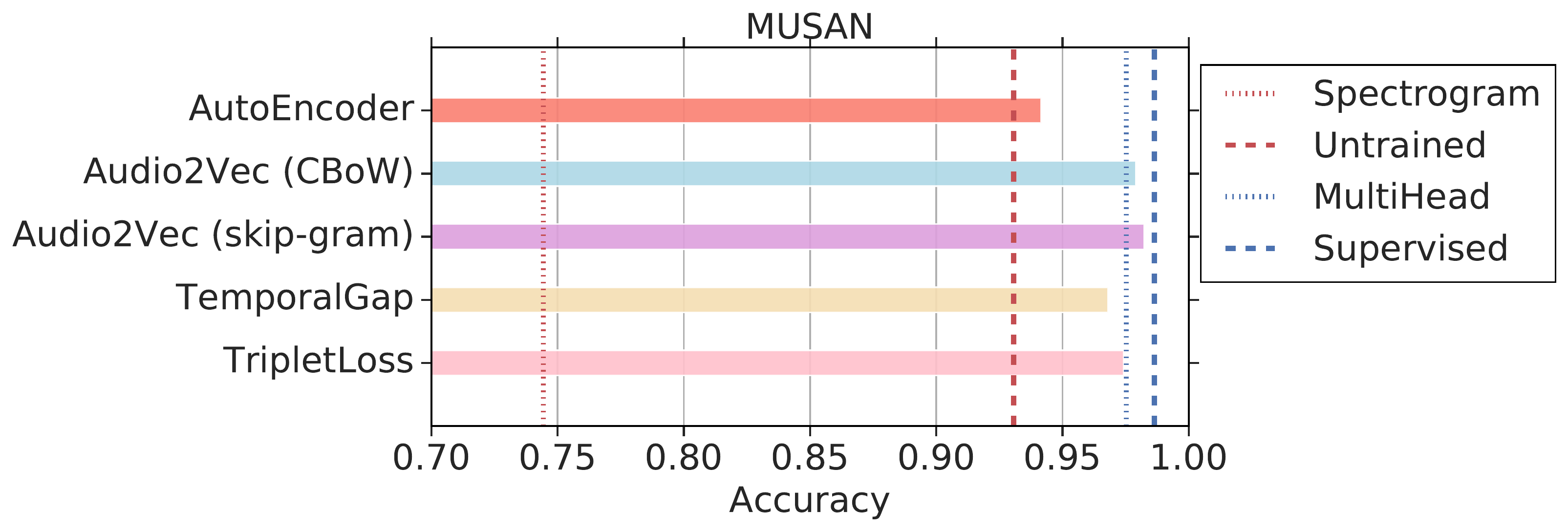}
        \caption{}
        \label{fig:fig_MUSAN}
    \end{subfigure}
    \caption{(a): Training loss for the {\audiotovec} (skip-gram) model and corresponding accuracy on the {\MUSAN} downstream dataset. (b) Accuracy on {\MUSAN} datasets for all models under evaluation.}\label{fig:MUSAN_results}
\end{figure}

\begin{table}
  \begin{center}
    \caption{Accuracy on downstream tasks (and fraction of accuracy recovered wrt. baselines). Downstream tasks: SPC: ({\speechcommands}), LSP: ({\librispeech}), TUT: {\TUT}, MUS: {\MUSAN}, BSD: {\birdsong}, LID: {\langid}. In \textbf{bold} the highest accuracy attained by self-supervised models for each task.}
    \label{tab:accuracy}
    
\begin{tabular}{rcccccc}
\hline
 model  & SPC  & LID  & LSP  & MUS  & TUT  & BSD \\
\hline
\rowcolor{LightRed}
  \texttt{Spectrogram} & 0.16 $\pm$ .01& 0.28 $\pm$ .04& 0.97 $\pm$ .01& 0.74 $\pm$ .01& 0.36 $\pm$ .03& 0.65 $\pm$ .02\\
\rowcolor{LightRed}
 & {\small {\color{darkgray} (+0\%) }}& {\small {\color{darkgray} (+0\%) }}& {\small {\color{darkgray} (+0\%) }}& {\small {\color{darkgray} (+0\%) }}& {\small {\color{darkgray} (+0\%) }}& {\small {\color{darkgray} (+0\%) }}\\
 \rowcolor{LightRed}
  \texttt{Untrained} & 0.16 $\pm$ .01& 0.48 $\pm$ .04& 0.54 $\pm$ .02& 0.93 $\pm$ .00& 0.57 $\pm$ .03& 0.70 $\pm$ .02\\
  \rowcolor{LightRed}
 & {\small {\color{darkgray} (-1\%) }}& {\small {\color{darkgray} (+33\%) }}& {\small {\color{darkgray} (-1338\%) }}& {\small {\color{darkgray} (+77\%) }}& {\small {\color{darkgray} (+35\%) }}& {\small {\color{darkgray} (+31\%) }}\\
  \texttt{AutoEncoder} & 0.28 $\pm$ .01& \textbf{0.64} $\pm$ .04& 0.99 $\pm$ .00& 0.94 $\pm$ .00& 0.59 $\pm$ .03& 0.69 $\pm$ .02\\
 & {\small {\color{darkgray} (+21\%) }}& {\small {\color{darkgray} (+56\%) }}& {\small {\color{darkgray} (+55\%) }}& {\small {\color{darkgray} (+81\%) }}& {\small {\color{darkgray} (+38\%) }}& {\small {\color{darkgray} (+27\%) }}\\
  \texttt{A2V(CBoW)} & \textbf{0.30} $\pm$ .01& 0.57 $\pm$ .04& 0.99 $\pm$ .00& \textbf{0.98} $\pm$ .00& 0.66 $\pm$ .03& 0.71 $\pm$ .01\\
 & {\small {\color{darkgray} (+23\%) }}& {\small {\color{darkgray} (+47\%) }}& {\small {\color{darkgray} (+82\%) }}& {\small {\color{darkgray} (+97\%) }}& {\small {\color{darkgray} (+50\%) }}& {\small {\color{darkgray} (+40\%) }}\\
  \texttt{A2V(SG)} & 0.28 $\pm$ .01& 0.55 $\pm$ .04& \textbf{1.00} $\pm$ .00& \textbf{0.98} $\pm$ .00& 0.67 $\pm$ .03& 0.69 $\pm$ .02\\
 & {\small {\color{darkgray} (+21\%) }}& {\small {\color{darkgray} (+44\%) }}& {\small {\color{darkgray} (+85\%) }}& {\small {\color{darkgray} (+98\%) }}& {\small {\color{darkgray} (+52\%) }}& {\small {\color{darkgray} (+28\%) }}\\
  \texttt{TemporalGap} & 0.23 $\pm$ .01& 0.45 $\pm$ .04& 0.97 $\pm$ .01& 0.97 $\pm$ .00& 0.63 $\pm$ .03& 0.71 $\pm$ .01\\
 & {\small {\color{darkgray} (+12\%) }}& {\small {\color{darkgray} (+27\%) }}& {\small {\color{darkgray} (+11\%) }}& {\small {\color{darkgray} (+92\%) }}& {\small {\color{darkgray} (+44\%) }}& {\small {\color{darkgray} (+44\%) }}\\
  \texttt{TripletLoss} & 0.18 $\pm$ .01& 0.62 $\pm$ .04& \textbf{1.00} $\pm$ .00& 0.97 $\pm$ .00& \textbf{0.73} $\pm$ .03& \textbf{0.73} $\pm$ .01\\
 & {\small {\color{darkgray} (+3\%) }}& {\small {\color{darkgray} (+55\%) }}& {\small {\color{darkgray} (+96\%) }}& {\small {\color{darkgray} (+95\%) }}& {\small {\color{darkgray} (+61\%) }}& {\small {\color{darkgray} (+55\%) }}\\
 \rowcolor{LightCyan}
  \texttt{MultiHead} & 0.72 $\pm$ .01& 0.82 $\pm$ .03& 1.00 $\pm$ .00& 0.98 $\pm$ .00& 0.94 $\pm$ .02& 0.78 $\pm$ .01\\
   \rowcolor{LightCyan}
 & {\small {\color{darkgray} (+95\%) }}& {\small {\color{darkgray} (+88\%) }}& {\small {\color{darkgray} (+99\%) }}& {\small {\color{darkgray} (+95\%) }}& {\small {\color{darkgray} (+96\%) }}& {\small {\color{darkgray} (+90\%) }}\\
  \rowcolor{LightCyan}
  \texttt{Supervised} & 0.75 $\pm$ .01& 0.90 $\pm$ .03& 1.00 $\pm$ .00& 0.99 $\pm$ .00& 0.97 $\pm$ .01& 0.79 $\pm$ .01\\
   \rowcolor{LightCyan}
 & {\small {\color{darkgray} (+100\%) }}& {\small {\color{darkgray} (+100\%) }}& {\small {\color{darkgray} (+100\%) }}& {\small {\color{darkgray} (+100\%) }}& {\small {\color{darkgray} (+100\%) }}& {\small {\color{darkgray} (+100\%) }}\\
\hline
\end{tabular}

  \end{center}
  
\end{table}

\textbf{Main results}: In our results we report the prediction accuracy on the eval set of each of the six datasets. 
During training we monitor both the loss of the self-supervised task as well as the accuracy on each of the downstream tasks. As an illustration, Figure~\ref{fig:self_task_audio_to_vec_skip_gram_MUSAN} shows that the accuracy of the {\MUSAN} downstream task increases as the reconstruction loss of {\audiotovec} (skip-gram) decreases, and both tend to saturate after approximately 300k iterations. 
For the same dataset, Figure~\ref{fig:fig_MUSAN} shows that all self-supervised methods attain a level of accuracy that is in-between the baselines and the fully supervised benchmarks, with {\audiotovec} (skip-gram) outperforming the other models on this task. We repeated the evaluation on all downstream tasks and show the results in Table~\ref{tab:accuracy}. 
We report the level of accuracy, with 95\% confidence intervals capturing the uncertainty due to the finite size of the evaluation datasets. In brackets we also report the accuracy normalized between 0\% ({\spectrogram}) and 100\% ({\singlehead}). We observe that the proposed self-supervised learning models are able to recover between 11\% and 98\% of the accuracy of the {\singlehead} model. Generally, {\audiotovec} (skip-gram) and {\tripletloss} seem to outperform other self-supervised models. The best results are obtained on {\MUSAN} and {\librispeech}, presumably because these tasks require to capture relatively stationary spectral characteristics of the inputs. Conversely, all self-supervised models achieve relatively poor performance on the {\speechcommands} dataset. This might be explained by the fact that for this dataset it is particularly important to recognize the non-stationary variation of the spectral features along the temporal dimension, which does not seem to be captured by the embeddings generated by the self-supervised models. 
Note that the different self-supervised models might be capturing different characteristics of the underlying audio data. Therefore, there might be the possibility of merging the different representations, as recently proposed in \cite{Pascual2019} for the case of speech embeddings. 

We repeated a similar evaluation working directly in the embedding space, by training a simple $k$-nearest neighbour classifier ($k$=10) on each dataset. More specifically, we extracted 975ms samples at random from the original audio clips (10000 samples for training and 2000 samples for evaluation, for each dataset), and mapped each sample to a 128-dimensional embedding. The classifier computes Euclidean distances directly in the embedding space. For the {\spectrogram} baseline, we first perform dimensionality reduction by applying a random projection matrix sampled from a Gaussian distribution to map the flattened $96\times 64$ spectrogram to a 128-dimensional space. Table~\ref{tab:knn_accuracy} reports the results, showing that also in this case the proposed self-supervised models recover between 10\% and 99\% of the accuracy of the {\singlehead} model. This demonstrates that salient representations of the underlying audio data are indeed captured directly in the embedding space. 

\begin{table}
  \begin{center}
    \caption{Accuracy on kNN classification (and fraction of accuracy recovered wrt. baselines).}
    \label{tab:knn_accuracy}
    
 \begin{tabular}{rcccccc}
 \hline
  model  & SPC  & LID  & LSP  & MUS  & TUT  & BSD \\
 \hline
 \hline
  \rowcolor{LightRed}
   \texttt{Spectrogram} & 0.02 $\pm$ .01& 0.39 $\pm$ .02& 0.00 $\pm$ .00& 0.10 $\pm$ .01& 0.11 $\pm$ .01& 0.49 $\pm$ .02\\
     \rowcolor{LightRed}
  & {\small {\color{darkgray} (+0\%) }}& {\small {\color{darkgray} (+0\%) }}& {\small {\color{darkgray} (+0\%) }}& {\small {\color{darkgray} (+0\%) }}& {\small {\color{darkgray} (+0\%) }}& {\small {\color{darkgray} (+0\%) }}\\
    \rowcolor{LightRed}
   \texttt{Untrained} & 0.08 $\pm$ .01& 0.38 $\pm$ .02& 0.04 $\pm$ .01& 0.87 $\pm$ .01& 0.41 $\pm$ .02& 0.68 $\pm$ .02\\
     \rowcolor{LightRed}
  & {\small {\color{darkgray} (+9\%) }}& {\small {\color{darkgray} (-3\%) }}& {\small {\color{darkgray} (+3\%) }}& {\small {\color{darkgray} (+88\%) }}& {\small {\color{darkgray} (+41\%) }}& {\small {\color{darkgray} (+77\%) }}\\
   \texttt{AutoEncoder} & \textbf{0.24} $\pm$ .02& \textbf{0.44} $\pm$ .02& 0.03 $\pm$ .01& 0.68 $\pm$ .02& 0.52 $\pm$ .02& 0.67 $\pm$ .02\\
  & {\small {\color{darkgray} (+30\%) }}& {\small {\color{darkgray} (+20\%) }}& {\small {\color{darkgray} (+3\%) }}& {\small {\color{darkgray} (+67\%) }}& {\small {\color{darkgray} (+55\%) }}& {\small {\color{darkgray} (+70\%) }}\\
   \texttt{A2V(CBoW)} & 0.14 $\pm$ .02& 0.43 $\pm$ .02& 0.10 $\pm$ .01& 0.94 $\pm$ .01& 0.52 $\pm$ .02& 0.69 $\pm$ .02\\
  & {\small {\color{darkgray} (+17\%) }}& {\small {\color{darkgray} (+16\%) }}& {\small {\color{darkgray} (+10\%) }}& {\small {\color{darkgray} (+96\%) }}& {\small {\color{darkgray} (+55\%) }}& {\small {\color{darkgray} (+81\%) }}\\
   \texttt{A2V(SG)} & 0.12 $\pm$ .01& 0.43 $\pm$ .02& 0.26 $\pm$ .02& \textbf{0.96} $\pm$ .01& 0.60 $\pm$ .02& 0.70 $\pm$ .02\\
  & {\small {\color{darkgray} (+14\%) }}& {\small {\color{darkgray} (+15\%) }}& {\small {\color{darkgray} (+27\%) }}& {\small {\color{darkgray} (+99\%) }}& {\small {\color{darkgray} (+67\%) }}& {\small {\color{darkgray} (+84\%) }}\\
   \texttt{TemporalGap} & 0.10 $\pm$ .01& 0.37 $\pm$ .02& 0.35 $\pm$ .02& 0.92 $\pm$ .01& 0.55 $\pm$ .02& 0.70 $\pm$ .02\\
  & {\small {\color{darkgray} (+11\%) }}& {\small {\color{darkgray} (-10\%) }}& {\small {\color{darkgray} (+36\%) }}& {\small {\color{darkgray} (+93\%) }}& {\small {\color{darkgray} (+60\%) }}& {\small {\color{darkgray} (+84\%) }}\\
   \texttt{TripletLoss} & 0.09 $\pm$ .01& 0.25 $\pm$ .02& \textbf{0.69} $\pm$ .02& \textbf{0.96} $\pm$ .01& \textbf{0.70} $\pm$ .02& \textbf{0.72} $\pm$ .02\\
  & {\small {\color{darkgray} (+10\%) }}& {\small {\color{darkgray} (-59\%) }}& {\small {\color{darkgray} (+71\%) }}& {\small {\color{darkgray} (+99\%) }}& {\small {\color{darkgray} (+80\%) }}& {\small {\color{darkgray} (+91\%) }}\\
   \rowcolor{LightCyan}
   \texttt{MultiHead} & 0.69 $\pm$ .02& 0.52 $\pm$ .02& 0.86 $\pm$ .02& 0.95 $\pm$ .01& 0.75 $\pm$ .02& 0.75 $\pm$ .02\\
    \rowcolor{LightCyan}
  & {\small {\color{darkgray} (+91\%) }}& {\small {\color{darkgray} (+52\%) }}& {\small {\color{darkgray} (+89\%) }}& {\small {\color{darkgray} (+97\%) }}& {\small {\color{darkgray} (+88\%) }}& {\small {\color{darkgray} (+102\%) }}\\
   \rowcolor{LightCyan}
   \texttt{Supervised} & 0.76 $\pm$ .02& 0.63 $\pm$ .02& 0.97 $\pm$ .01& 0.97 $\pm$ .01& 0.84 $\pm$ .02& 0.74 $\pm$ .02\\
    \rowcolor{LightCyan}
  & {\small {\color{darkgray} (+100\%) }}& {\small {\color{darkgray} (+100\%) }}& {\small {\color{darkgray} (+100\%) }}& {\small {\color{darkgray} (+100\%) }}& {\small {\color{darkgray} (+100\%) }}& {\small {\color{darkgray} (+100\%) }}\\
 \hline
 \end{tabular}

  \end{center}
\end{table}

\textbf{Impact of training dataset}: All the results reported so far use the {\audioset} dataset to train the self-supervised models. {\audioset} contains a wide variety of audio clips, including music, speech, ambient noise, acoustic events, etc.
In order to evaluate the impact of the choice of the dataset, we repeated self-supervised training using {\librispeech} (discarding the speaker labels). We chose {\librispeech} because the original samples are sufficiently long to support our self-learning tasks and because it contains audio of different content than {\audioset} (i.e., speech only). Table~\ref{tab:training_on_librispeech} reports how the evaluation results shown in Table~\ref{tab:accuracy} change when training the self-supervised models on {\librispeech} instead of {\audioset}. In most cases, we observe a decrease in the level of accuracy on downstream tasks, especially for {\temporalgap} and {\tripletloss}, suggesting that a richer content variety in the training set is preferable when learning general-purpose audio representations.

\begin{table}
  \begin{center}
  \caption{Accuracy obtained when training self-supervised models on {\librispeech} (and the relative difference with respect to training on {\audioset}). Red indicates a decrease in the level of accuracy.}
  \label{tab:training_on_librispeech}
\begin{tabular}{rcccccc}
\hline
 model  & SPC  & LID  & LSP  & MUS  & TUT  & BSD \\
\hline
  \texttt{AutoEncoder} & 0.27 & 0.65 & 0.96 & 0.87 & 0.56 & 0.67 \\
  & {\small {\color{red} (-3\%) }}& {\small {\color{darkgray} (+1\%) }}& {\small {\color{red} (-3\%) }}& {\small {\color{red} (-7\%) }}& {\small {\color{red} (-5\%) }}& {\small {\color{red} (-2\%) }}\\
  \texttt{A2V(CBoW)} & \textbf{0.26} & 0.55 & 0.99 & 0.96 & 0.65 & 0.70 \\
  & {\small {\color{red} (-13\%) }}& {\small {\color{red} (-3\%) }}& {\small {\color{darkgray} (+0\%) }}& {\small {\color{red} (-2\%) }}& {\small {\color{red} (-1\%) }}& {\small {\color{red} (-1\%) }}\\
  \texttt{A2V(SG)} & 0.23 & \textbf{0.65} & 0.99 & \textbf{0.97} & \textbf{0.66} & \textbf{0.71} \\
  & {\small {\color{red} (-17\%) }}& {\small {\color{darkgray} (+18\%) }}& {\small {\color{red} (-1\%) }}& {\small {\color{red} (-1\%) }}& {\small {\color{red} (-1\%) }}& {\small {\color{darkgray} (+2\%) }}\\
  \texttt{TemporalGap} & 0.18 & 0.55 & 0.93 & 0.94 & 0.59 & 0.64 \\
  & {\small {\color{red} (-21\%) }}& {\small {\color{darkgray} (+22\%) }}& {\small {\color{red} (-4\%) }}& {\small {\color{red} (-3\%) }}& {\small {\color{red} (-6\%) }}& {\small {\color{red} (-9\%) }}\\
  \texttt{TripletLoss} & 0.10 & 0.34 & \textbf{1.00} & 0.93 & 0.56 & 0.65 \\
  & {\small {\color{red} (-44\%) }}& {\small {\color{red} (-45\%) }}& {\small {\color{darkgray} (+0\%) }}& {\small {\color{red} (-4\%) }}& {\small {\color{red} (-23\%) }}& {\small {\color{red} (-10\%) }}\\
\hline
\end{tabular}
  \end{center}
\end{table}

\begin{figure}
    \centering
    \begin{subfigure}[b]{0.48\textwidth}        \includegraphics[width=\textwidth]{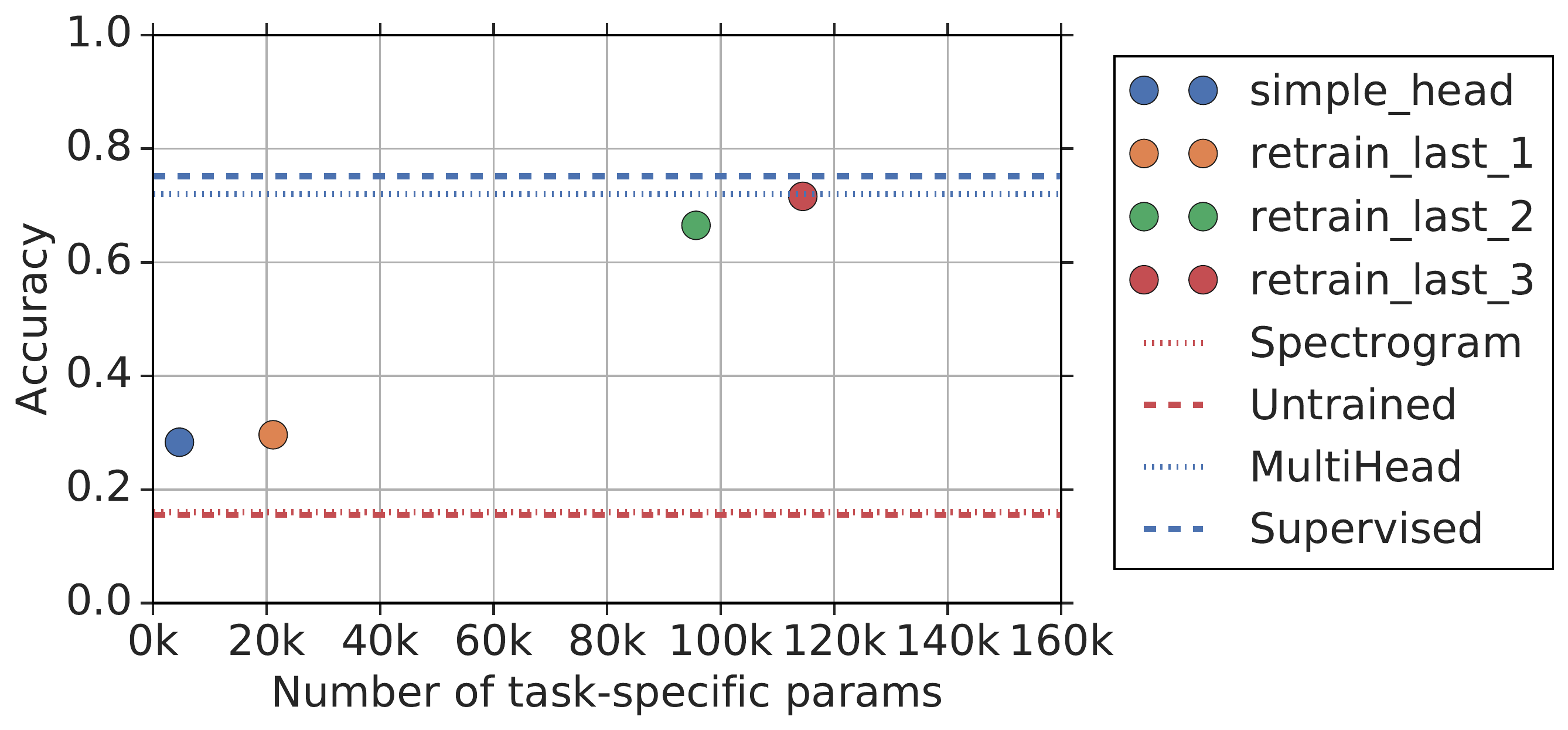}
        \caption{{\speechcommands}}
        \label{retrain_head_speech_commands_v2}
    \end{subfigure}
    ~ 
    \begin{subfigure}[b]{0.48\textwidth}
        \includegraphics[width=\textwidth]{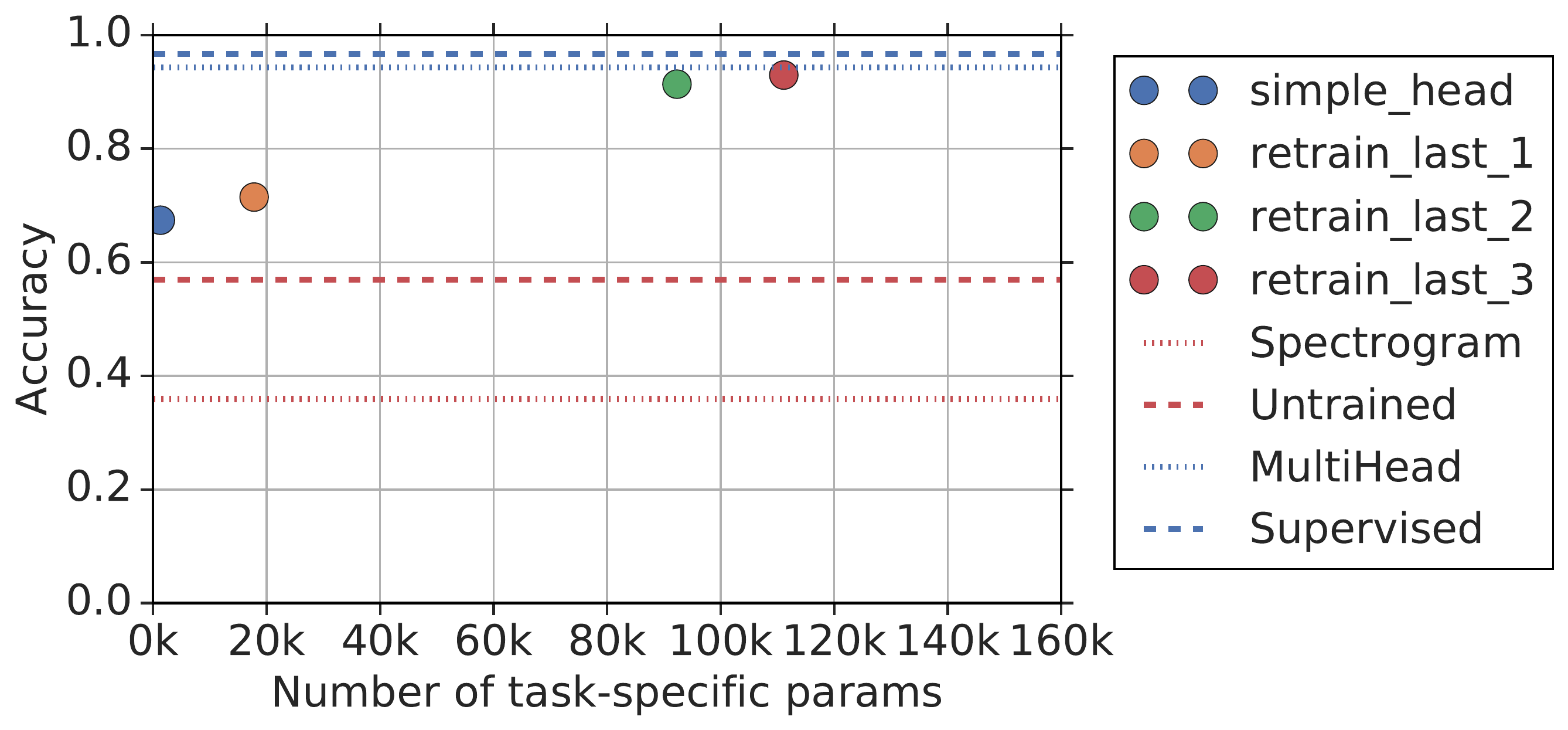}
        \caption{{\TUT}}
        \label{retrain_head_TUT-urban-acoustic-scenes-2018}
    \end{subfigure}
    \caption{Accuracy obtained when retraining the last layers of the {\audiotovec} (skip-gram) encoder.}\label{fig:retrain_head}
\end{figure}

\textbf{Encoder fine-tuning}: So far we considered the case in which the encoder is shared completely across different tasks, and only the last layer is allowed to learn task-specific parameters. It is interesting to observe what happens when we relax this assumption, allowing to retrain one (or more) of the deepest layers of the encoder. 
Figure~\ref{fig:retrain_head} shows the trade-off between the level of accuracy and the number of task specific parameters for two datasets, {\speechcommands} and {\TUT}, for which {\audiotovec} (skip-gram) was able to only partially bridge the accuracy gap with respect to the {\singlehead} model. The left-most (blue) point corresponds to the accuracy already reported in Table~\ref{tab:accuracy}. Note that in this case the number of task-specific parameters is equal to $128 \times C$, where $C$ is the number of classes (equal to, respectively, 35 and 10 for these datasets). The second (orange) point from the left corresponds to retraining the fully-connected layer, while the remaining points correspond to retraining the until the fifth and fourth convolutional layers included. Generally, retraining the last two layers is needed to recover most of the accuracy of the fully supervised model. Note that, although the last two layers account for approximately 80\% of the parameters, they only contribute to 20\% of the FLOPs, and this is particularly useful when deploying on mobile devices.

\textbf{Impact of encoder architecture size}: Although the focus of this paper is on encoder architectures that can be deployed on mobile devices, the proposed self-supervised methods are general and they can be applied also to larger models. Therefore, we repeated our evaluation by increasing the size of the encoder architecture described in Table~\ref{tab:encoder_architecture}. Namely, we increased the number of channels in each convolutional layer by a factor of 4, and we increased the number of outputs in the last fully connected layer to obtain 256-dimensional embeddings. Table~\ref{tab:large_encoder} shows that the accuracy on downstream tasks increases, and {\audiotovec} (skip-gram) achieves the highest accuracy on almost all datasets.

\begin{table}[t]
  \begin{center}

  \caption{Accuracy obtained when using a larger encoder architecture (relative change wrt. Table~\ref{tab:accuracy}). }
  \label{tab:large_encoder}

\begin{tabular}{rcccccc}
\hline
 model  & SPC  & LID  & LSP  & MUS  & TUT  & BSD \\
\hline
\hline
  \texttt{AutoEncoder} & 0.35 & 0.62 & \textbf{1.00} & 0.96 & 0.65 & 0.70 \\
  & {\small {\color{darkgray} (+24\%) }}& {\small {\color{darkgray} (-3\%) }}& {\small {\color{darkgray} (+1\%) }}& {\small {\color{darkgray} (+2\%) }}& {\small {\color{darkgray} (+10\%) }}& {\small {\color{darkgray} (+1\%) }}\\
  \texttt{A2V(SG)} & \textbf{0.46} & \textbf{0.81} & \textbf{1.00} & \textbf{0.99} & 0.78 & \textbf{0.76} \\
  & {\small {\color{darkgray} (+64\%) }}& {\small {\color{darkgray} (+47\%) }}& {\small {\color{darkgray} (+0\%) }}& {\small {\color{darkgray} (+1\%) }}& {\small {\color{darkgray} (+16\%) }}& {\small {\color{darkgray} (+10\%) }}\\
  \texttt{TemporalGap} & 0.37 & 0.77 & \textbf{1.00} & 0.98 & 0.73 & 0.74 \\
  & {\small {\color{darkgray} (+60\%) }}& {\small {\color{darkgray} (+71\%) }}& {\small {\color{darkgray} (+3\%) }}& {\small {\color{darkgray} (+1\%) }}& {\small {\color{darkgray} (+15\%) }}& {\small {\color{darkgray} (+4\%) }}\\
  \texttt{TripletLoss} & 0.30 & 0.73 & \textbf{1.00} & \textbf{0.99} & \textbf{0.81} & 0.76 \\
  & {\small {\color{darkgray} (+66\%) }}& {\small {\color{darkgray} (+17\%) }}& {\small {\color{darkgray} (+0\%) }}& {\small {\color{darkgray} (+2\%) }}& {\small {\color{darkgray} (+10\%) }}& {\small {\color{darkgray} (+4\%) }}\\
\hline
\end{tabular}
  \end{center}
\end{table}


\section{Conclusion}\label{sec:conclusions}
In this paper we present self-supervised learning methods that exploit the temporal context in audio clips. Our results show that both {\audiotovec} and {\temporalgap} are able to produce representations that can be re-used for different downstream tasks, without having access to labelled datasets during training. We based our experiments on small encoder architectures, which can be potentially deployed on mobile devices. This is motivated by the fact that, in our future work, we will investigate training self-supervised models directly on device in a distributed fashion, by taking advantage of federated learning. Another interesting direction is merging representations learned by different self-supervised models, as recently proposed in \cite{Pascual2019} for the case of speech embeddings.




{\small
\bibliographystyle{plainnat}
\bibliography{references}
}

\end{document}